\documentclass[aps,eqsecnum,nofootinbib,superscriptaddress,showpacs,twocolumn]{revtex4}
\usepackage{amsmath,amsfonts,amssymb,bm}
\usepackage[dvips]{graphicx}
\usepackage{color}
\usepackage{comment}
\def\be{\begin{eqnarray}}
\def\ee{\end{eqnarray}}
\def\bec{\begin{center}}
\def\eec{\end{center}}

\def\p{\partial}
\bmdefine{\bmj}{\bm{j}}
\bmdefine{\bmk}{\bm{k}}
\bmdefine{\bmx}{\bm{x}}
\bmdefine{\bmA}{\bm{A}}
\bmdefine{\bmD}{\bm{D}}
\bmdefine{\bmF}{\bm{F}}
\newcommand{\calE}{\mathcal{E}}
\newcommand{\calJ}{\mathcal{J}}
\newcommand{\calK}{\mathcal{K}}
\newcommand{\calM}{\mathcal{M}}
\newcommand{\calT}{\mathcal{T}}
\newcommand{\Exp}[1]{\left\langle~#1~\right\rangle}
\begin{document}
\title{Inhomogeneous charged black hole solutions in asymptotically anti-de Sitter spacetime}
\author{Kengo Maeda}
\email{maeda302@sic.shibaura-it.ac.jp}
\affiliation{Faculty of Engineering,
Shibaura Institute of Technology, Saitama, 330-8570, Japan}
\author{Takashi Okamura}
\email{tokamura@kwansei.ac.jp}
\affiliation{Department of Physics, Kwansei Gakuin University,
Sanda, 669-1337, Japan}
\author{Jun-ichirou Koga}
\email{koga@waseda.jp}
\affiliation{Research Institute for Science and Engineering, 
Waseda University, Shinjuku, Tokyo 169-8555, Japan}
\date{\today}
\begin{abstract}
We investigate static inhomogeneous charged planar black hole solutions
of the Einstein-Maxwell system in an asymptotically anti-de Sitter spacetime.
Within the framework of linear perturbations, the solutions are 
numerically and analytically constructed from the Reissner-Nordstr\"{o}m-AdS
black hole solution. 
The perturbation analysis predicts that the Cauchy horizon
always disappears for any wavelength perturbation, supporting the strong 
cosmic censorship conjecture. 
For extremal black holes, we analytically show that
an observer freely falling into the black hole feels infinite tidal force
at the horizon for any long wavelength perturbation,
even though the Kretschmann scalar curvature invariant remains small. 
\end{abstract}

\maketitle

\section{Introduction}\label{sec:intro}
Recently, much attention has been paid to the investigation of 
black holes in an asymptotically anti de Sitter~(AdS) spacetime.  
According to the AdS/CFT duality~\cite{Maldacena:1997re}, 
the gravitational theory in a black hole spacetime is dual to 
a strongly coupled gauge theory at finite temperature.  
As a fascinating application of the duality, a holographic model of 
superconductors has been constructed from black hole 
solutions with charged scalar hair~\cite{Hartnoll:2008kx}. 
This indicates that there is a variety of black hole solutions 
in an asymptotically AdS spacetime, which could be useful to understand 
strongly correlated condensed matter physics. 

Even in the Einstein-Maxwell system, many black hole solutions with 
various topologies can be constructed in an asymptotically 
AdS spacetime~\cite{Lemos1,Lemos2, Lemos3, Huang,Aminneborg,brill,Mann}, 
while a uniqueness theorem has been established for the static black  
hole solution without charge by restricting the topology to $S^2$~\cite{Anderson}. 
The planar Reissner-Nordstr\"{o}m AdS solution with $R^2$ topology is one of 
the black hole solutions~\cite{Huang}.  It attracts much attention as a holographic 
model of strongly correlated condensed matter systems because the dual theory 
lives in a flat 2+1 dimensional spacetime. 

When we apply the AdS/CFT duality to condensed matter systems, it is interesting   
to incorporate the lattice structure into the boundary theory. 
As it is typical in condensed matter physics, the lattice structure  
induces a periodic inhomogeneous electric potential.
In the holographic theory, such inhomogeneity in the boundary theory  
corresponds to that of the gauge field in the bulk theory.
So, in this paper, we construct inhomogeneous charged static black hole 
solutions, by perturbing the planar Reissner-Nordstr\"{o}m AdS solution, 
and investigate their geometrical properties. 

On the other hand, from the perspective of General Relativity, one of the issues 
in the bulk spacetime is whether a naked singularity appears. According to 
the (strong) cosmic censorship hypothesis~\cite{penrose}, any singularity should be hidden inside 
the event horizon and the geometry cannot be extended beyond the Cauchy horizon. However, in the 
case of the Reissner-Nordstr\"{o}m AdS solution, there exists a region where a naked singularity 
can be observed behind the regular Cauchy horizon.  
Even though the Cauchy horizon is unstable against dynamical perturbations 
due to the infinite blueshift~\cite{ChandHartle}, the possibility to observe 
a singularity should still remain because the resulting singularity would be 
a null weak singularity~\cite{bradysmith}. 
In the context of the AdS/CFT duality, the hypothesis was also argued 
in~\cite{LeviRoss2003,Balasubramanian2004}. 

Another issue regarding the cosmic censorship hypothesis is 
whether the zero temperature black hole solutions can have a regular event horizon. 
In a class of extremal black holes in string theory, it has been shown that the event horizon 
cannot be smooth even though the Kretschmann scalar curvature invariant 
$R_{\mu\nu\alpha\beta}R^{\mu\nu\alpha\beta}$ is small there~\cite{horowitzross1997}. 
Recently, a similar phenomenon was observed in the Lifshitz spacetime~\cite{MannCop}. 
For the AdS black hole solutions with charged scalar hair, it has been shown that 
the event horizon cannot be regular in the extremal limit~\cite{Fernandez}. 
So, the extremal Reissner-Nordstr\"{o}m AdS solution with a regular event horizon 
seems to be exceptional. 

Therefore, in this paper, we will particularly investigate the geometry inside 
the event horizon of our inhomogeneous charged black hole solutions,
both numerically and analytically. Being quite different from the unperturbed 
Reissner-Nordstr\"{o}m AdS solution, the Kretschmann scalar curvature invariant 
becomes infinitely large toward the Cauchy horizon for any wavelength perturbations. 
In the extremal case, we show that the black hole solutions with long wavelength 
inhomogeneity cannot have a regular event horizon even though 
the Kretschmann scalar curvature invariant is small there. This is because 
the tidal force which an observer freely falling into the black hole experiences 
diverges towards the horizon. This tidal force is strong in the sense that the 
shear of any timelike geodesic congruence diverges infinitely. 
Hence, smooth extension of the geometry beyond the event horizon is impossible. 

In the next section, we derive the static perturbed equations of the 
planar Reissner-Nordstr\"{o}m-AdS black hole solution. 
In Sec. III we numerically and analytically construct the non-extremal solutions 
and observe that the curvature blows up towards the Cauchy horizon for any 
wavelength perturbation. In Sec.~IV, we analyze the extremal solutions. 
Conclusion and discussions are devoted in Sec.~V.

\section{Static perturbations of Reissner-Nordstr\"{o}m-AdS black hole}
\label{sec:solution}
We consider the four-dimensional Einstein-Maxwell system in an asymptotically 
anti-de Sitter spacetime with
the action%
\footnote{Here, we set $16\pi G=1$.}
\begin{align}
  & S = \int d^4x~\sqrt{-g}\, \left( R + \frac{6}{L^2}
  - \frac{1}{4}\, F_{\mu\nu} F^{\mu\nu} \right),
\label{action}
\end{align} 
where $L$ is the AdS curvature radius
and $F_{\mu\nu} = 2\, \p_{[\mu} A_{\nu]}$.
The field equations are 
\begin{subequations}
\label{field-eq}
\begin{align}
  & G_{\mu\nu}
  = \frac{3}{L^2}\, g_{\mu\nu}
  + \frac{1}{2} \left( F_{\mu\alpha}{F_\nu}^\alpha
    - \frac{1}{4}\, g_{\mu\nu}\,F^2 \right),
\label{field-eq_GW} \\
  & \sqrt{-g}~\nabla_\nu F^{\mu\nu}
  = \p_\nu ( \sqrt{-g}F^{\mu\nu} ) = 0.
\label{field-eq_EM}
\end{align}
\end{subequations}
The unperturbed plane-symmetric static black hole solution is the 
Reissner-Nordstr\"{o}m-AdS black hole solution:
\begin{subequations}
\label{background-sol}
\begin{align}
  & ds^2
  = \bar{g}_{\mu\nu}(u)\, dx^\mu dx^\nu
\nonumber \\
  &= \frac{L^2}{u^2} \left( - g(u)\, dt^2 + \frac{du^2}{g(u)}
  + dx^2 + dy^2 \right),
\label{eq:bg_metric} \\
  & g(u) := 1 - (1 + c) u^3 + c u^4,
\label{eq:def-g} \\
  & \bar{A}_\mu dx^\mu=\bar{A}_t(u) dt = \frac{2\, Q L^2}{r_+^2}\, (1 - u)\, dt,
  \hspace{0.5truecm}
  c := \frac{Q^2 L^2}{r_+^4},
\label{eq:bg-A}
\end{align}
\end{subequations}
where $Q$ and $r_+$ are the charge density
and the radius of the black hole.
The horizon and the spatial infinity are located at $u=1$ and $u=0$,
respectively.
Note that $c\le 3$ for all the black hole solutions
and the upper bound corresponds to the extremal black hole. 
According to the AdS/CFT duality,
the chemical potential $\mu$ is given by
$\mu=\lim_{u\to 0}\bar{A}_t(u)=2Q L^2/r_+^2$~\cite{Chamblin99}.

We consider static linear perturbations of the solution
(\ref{background-sol}) by adding a small chemical potential
with sinusoidal fluctuation in the $x$-direction.
In the scalar-type static perturbations, we can set as $g_{xx}=g_{yy}$ 
and $g_{xu}=0$ by a suitable gauge choice~\cite{kodama-ishibashi2004}. 
Then, we take the metric ansatz 
\begin{align}
\label{metric-u}
   ds^2
  &= \frac{L^2}{u^2} \Big[ - g(u) ( 1 + \epsilon a(u) e^{iqx}) dt^2
  + \frac{1 + \epsilon b(u) e^{iqx}}{g(u)}\, du^2
\nonumber \\
  &+ (1 + 2 \epsilon F(u) e^{iqx}) (dx^2 + dy^2)\, \Big]~,
\end{align}
with the gauge field
\begin{align}
\label{def-Bt}
   A_\mu dx^\mu = A_t(u,x)dt
  = \left(\bar{A}_t + \frac{\epsilon r_+^2}{Q} B_t(u) e^{iqx}
  \right) dt.
\end{align}
It is noteworthy that the variables $a$, $b$, and $F$ agree with  
the gauge invariant quantities adopted in Ref.~\cite{kodama-ishibashi2004}
for the gauge choice (\ref{metric-u}). 

For the electromagnetic perturbations~(\ref{def-Bt}),
we immediately obtain 
\begin{align}
a(u)=-b(u) 
\end{align}
from the combination of $xx$ and $yy$ components of the 
field equations~(\ref{field-eq_GW}). 
Let us introduce a variable $Y$ as
\begin{align}
\label{def-Y}
& Y(u):=b(u)-2F(u)=-a(u)-2F(u).
\end{align}
In terms of the three variables, $F$, $Y$, and $B_t$,
the following four coupled differential equations are derived: 
\begin{subequations}
\label{pert-field1-eq}
\begin{align}
   0
  &= \dot{Y}
  + \left( \frac{\dot{g}(u)}{g(u)} - \frac{2}{u} \right) (2 F + Y)
  - \frac{2 u^2}{g(u)}\, B_t,
\label{eq-constraint1} \\
   0
  &=
  \frac{u^3}{g(u)} \left( \Dot{B}_t - \frac{2B_t}{u} \right)
  + u \left( \frac{2}{u} - \frac{\dot{g}(u)}{g(u)} \right) \dot{F}
\nonumber \\
  &
  + \frac{2}{u}\, F
  + \left( \frac{1}{u} - \frac{q^2\, u}{2\, g(u)} \right) Y~,
\label{Hamiltonian-constraint} \\
   0
  &= \ddot{F} - \frac{q^2}{g(u)} \left( F + \frac{Y}{2} \right),
\label{eq-F1} \\
   0
  &= \ddot{B}_t - 4 c \dot{F} - \frac{q^2}{g(u)}\, B_t,
\label{eq-Bt}
\end{align}
\end{subequations}
where a dot means the derivative with respect to $u$.
The first and second equations (\ref{eq-constraint1}), 
(\ref{Hamiltonian-constraint})
correspond to the momentum and the Hamiltonian constraint equations, 
respectively, when we foliate the spacetime by timelike hypersurfaces
homeomorphic to the AdS boundary. 
These equations are not independent of each other
because Eq.~(\ref{eq-Bt}) is derived from
Eqs.~(\ref{eq-constraint1})-(\ref{eq-F1}). Hence, 
Eqs.~(\ref{pert-field1-eq}) admit four linearly-independent 
mode solutions. 

To construct the solutions, we shall impose the following 
boundary conditions: 
\begin{itemize}
\item The spacetime is an asymptotically AdS spacetime.  
\item The solution is regular at the horizon in the 
sense that the Kretschmann scalar curvature invariant
$\calK := R_{\mu\nu\alpha\beta}R^{\mu\nu\alpha\beta}$ is 
bounded at the horizon, 
\begin{align}
\label{cond-curvature}
  \left|\, \calK(u=1)\, \right|
 = \left|\, R_{\mu\nu\alpha\beta}R^{\mu\nu\alpha\beta}(u=1)\, \right|
 < \infty~.
\end{align} 
\end{itemize}
At the horizon, the gauge invariant part 
$\delta \calK =: \delta\calK_q(u)\, e^{i q x}$
of the perturbation of the Kretschmann scalar curvature invariant
is expressed as%
\begin{align}
\label{curvature-behavior}
   \delta {\calK_q(1)}
  & \simeq \frac{4\epsilon}{L^4}
  \Big[ q^2 \{ (3 - 5c) Y + 6 (1 - c) F \}
\nonumber \\
  &- 2 (3 - 5c) \{ (3-c) \dot{F} - 2 B_t \} 
  \Big]
  + O(\epsilon^2)~,
\end{align}
for the gauge choice (\ref{metric-u}). 
Then, using Eqs.~(\ref{pert-field1-eq}), we can show that 
the condition~(\ref{cond-curvature}) is satisfied if and only if
the variables $F$, $Y$, and $B_t$ are finite at the horizon.
The asymptotic AdS boundary condition requires
$a(u=0) = b(u=0) = F(u=0)=0$. 
As easily checked from Eqs.~(\ref{pert-field1-eq}), $a(u=0) = b(u=0)=0$
if $F(u=0)=0$. Therefore, the boundary conditions  
can be described in terms of the variables $F$, $Y$, and $B_t$ as
\begin{subequations}
\label{eq:bc}
\begin{align}
  & F(u=0) = 0~,
\label{boundary-con} \\
  & |\, F(u=1)\, |,~~|\, Y(u=1)\, |,~~|\, B_t(u=1)\, | < \infty~.
\label{horizon-regularity}
\end{align}
\end{subequations}

As seen later, there are two mode solutions
satisfying the condition~(\ref{horizon-regularity}).
If a perturbed chemical potential $\delta \mu(x)=\delta \mu_q e^{iqx}$
is given at the boundary, a unique solution of Eqs.~(\ref{pert-field1-eq})
can be derived from the condition~(\ref{boundary-con}).
In other words, the normalization of the solution
is determined by the amplitude $\delta \mu_q$
of the inhomogeneous chemical potential.
For the discussions below, however,
we are concerned with linear perturbations,
and hence the amplitude of the perturbation can be re-normalized
at our disposal, and we will do so in what follows.

The condition~(\ref{boundary-con}) guarantees
that the dual theory lives in a flat $2+1$ dimensional spacetime
even though the bulk spacetime is inhomogeneous.
In the case of vacuum perturbations,
it is impossible to construct an inhomogeneous black hole solution
under the boundary conditions,
as the uniqueness theorem is established in the vacuum case
\cite{Anderson}
\footnote{An inhomogeneous black hole solution with $S^2$ horizon 
has been constructed in the vacuum case
when the perturbations asymptote to constant values~\cite{yoshino}.}.
In the following sections, we analytically and numerically 
construct inhomogeneous charged black hole solutions satisfying
the boundary conditions (\ref{eq:bc}). 

\section{Non-extremal solutions}
\subsection{Analytic solutions in the long wavelength limit}
We can construct analytically an inhomogeneous non-extremal
charged black hole solution of Eqs.~(\ref{pert-field1-eq})
in the long wavelength limit $q\to 0$.
In this subsection, we give the solution
as the first few terms in a power series in $q$.
Under the boundary condition~(\ref{boundary-con}),
we can read off from Eq.~(\ref{pert-field1-eq}) the behavior of the variables 
$F$, $Y$, and $B_t$ near the AdS boundary as
\begin{align}
\label{asym-behavior}
& F(u)\simeq u-\frac{q^2}{6}u^3, \qquad 
Y(u)\simeq -4u+O(u^3), \nonumber \\
& B_t(u)\simeq \gamma_1+\gamma_2\,u. 
\end{align}
Here, we normalized $F$ as $\lim_{u\to 0}F(u)/u=1$ for simplicity.

We can adopt Eqs.~(\ref{eq-constraint1}), (\ref{Hamiltonian-constraint}),
and (\ref{eq-F1}) as fundamental equations of the static perturbations
which admit four-independent solutions.
As easily checked, Eq.~(\ref{curvature-behavior}) diverges
at the horizon for the solutions that are non-analytic
in a neighborhood of the horizon $u=1$. %
So, the solutions satisfying
the regularity condition~(\ref{horizon-regularity}) are analytic 
in a neighborhood of the horizon. 
From the analyticity and Eq.~(\ref{eq-Bt}), 
we obtain   
\begin{align}
\label{gauge-horizon}
B_t(1)=0.  
\end{align}  

Let us expand the variables $F$, $Y$, and $B_t$ 
as a series in $q^2$: 
\begin{align}
   F(u)
  &= u + q^2\, F_1(u) + O (q^4)~, 
\nonumber \\
   B_t(u)
  &= B_{t0}(u) + q^2\, B_{t1}(u)+ O(q^4)~,
\label{expansion-inq} \\
   Y(u)
  &= Y_0(u) + q^2\, Y_1(u) + O(q^4)~. 
\nonumber
\end{align}
%

The zeroth and first order equations are 
\begin{subequations}
\label{expansion-zeroth}
\begin{align}
   0
  &= \dot{Y_0}
  + \left( \frac{\dot{g}(u)}{g(u)} - \frac{2}{u} \right) (2 F_0 + Y_0)
  - \frac{2 u^2}{g(u)}\, B_{t0},
\label{eq-Y0} \\
   0
  &=
  \frac{u^3}{g(u)} \left(\Dot{B}_{t0} - \frac{2B_{t0}}{u} \right)
  + u \left( \frac{2}{u} - \frac{\dot{g}(u)}{g(u)} \right) \dot{F}_0
\nonumber \\
  &
  + \frac{2\, F_0 + Y_0}{u}
  ~,
\label{Hamiltonian-constraint0}
\end{align}
\end{subequations}
%
and 
\begin{subequations}
\label{expansion-first}
\begin{align}
   0
  &= \dot{Y}_1
  + \left( \frac{\dot{g}(u)}{g(u)} - \frac{2}{u} \right) (2 F_1 + Y_1)
  - \frac{2 u^2}{g(u)}\, B_{t1},
\label{eq-Y1} \\
   0
  &=
  \frac{u^3}{g(u)} \left(\Dot{B}_{t1} - \frac{2B_{t1}}{u} \right)
  + u \left( \frac{2}{u} - \frac{\dot{g}(u)}{g(u)} \right) \dot{F}_1
\nonumber \\
  &
  + \frac{2\, F_1 + Y_1}{u} - \frac{u}{2\, g(u)}\, Y_0
  ~,
\label{Hamiltonian-constraint1} \\
   0
  &= \ddot{F}_1 - \frac{1}{g(u)} \left( F_0 + \frac{Y_0}{2} \right).
\label{eq-hatF0} 
\end{align}
\end{subequations}
%
The solution of Eqs.~(\ref{expansion-zeroth})
satisfying the boundary conditions~(\ref{horizon-regularity})
and (\ref{gauge-horizon}) is given by
\footnote{
Eq.(\ref{eq-Bt}) is useful to guess the solution (\ref{sol-B0}).
}
\begin{subequations}
\label{sol-zeroth}
\begin{align}
  & B_{t0}(u)
  = (1 - u)\, (3 + c - 2\, c\, u)
  ~,
\label{sol-B0} \\
  & Y_0(u)
  = - 2\, u\, \frac{ 2 + 2\, u - (1 + c)\, u^2}{1 + u + u^2 - c\, u^3}
  ~.
\label{sol-Y0}
\end{align}
\end{subequations}
%
%

By similar procedure, we can construct the first order solutions,
$F_1$, $B_{t1}$, and $Y_1$.
We give the explicit form in the Appendix B.
Thus, we obtain the analytic solution up to $O(q^2)$
which satisfy both the asymptotic boundary condition~(\ref{asym-behavior})
and the regularity conditions at the horizon,
(\ref{horizon-regularity}) and (\ref{gauge-horizon}).

\subsection{Numerical solutions and the curvature 
growth near the Cauchy horizon}
In this subsection we numerically construct inhomogeneous charged black hole solutions 
in an asymptotically AdS spacetime for various wavelengths. 
Eliminating $F(u)$ from Eqs.~(\ref{pert-field1-eq}), we obtain 
two coupled differential equations, 
\begin{subequations}
\begin{align}
& g(u)\left(\frac{2g(u)}{u}-\dot{g}(u)\right)\ddot{Y}(u)
\nonumber \\
&-\left(2\dot{g}^2(u)-\frac{2g(u)\dot{g}(u)}{u}
-2g\ddot{g}(u)+4cu^2g(u)\right)\dot{Y}(u) \nonumber \\
&-q^2\left(\frac{2g(u)}{u}-\dot{g}(u) \right)Y(u)-24B_t(u)=0, 
& \label{eq-Y} \\
& g(u)\{u\dot{g}(u)-2g(u)\}^2\ddot{B_t}(u) \nonumber \\
&-4cu^3g(u)(u\dot{g}(u)-2g(u)) 
\dot{B_t}(u) \nonumber \\
&+\{-4(q^2-6cu^2)g^2(u)-q^2u^2\dot{g}^2(u) \nonumber \\
&+4ug(u)[(q^2-4cu^2)\dot{g}(u)+cu(-12+u^2\ddot{g}(u))]\}
B_t(u) \nonumber \\
&+2cg^2(u)[2g(u)+u(u\ddot{g}(u)-2\dot{g}(u)-4cu^3)]\dot{Y}(u)
\nonumber \\
&+2cq^2ug(u)\{-2g(u)+u\dot{g}(u)\}Y(u)=0. 
& \label{eq-Bt1}
\end{align}
\label{two-coupled-eq_non-extreme}
\end{subequations}

As mentioned in the previous subsection, the solutions satisfying
the regularity conditions~(\ref{horizon-regularity})
and (\ref{gauge-horizon}) are analytic 
in a neighborhood of the horizon. 
By imposing analyticity for the variables $Y$ and $B_t$, we obtain 
$2\dot{Y}(1)=-q^2Y(1)/(3-c)$ from Eqs.~(\ref{two-coupled-eq_non-extreme}). 
For simplicity, we shall normalize $Y(1)$ as $Y(1)=3-c$ for the non-extremal 
black holes~(Recall $c<3$).  
Then, the only free parameter at the horizon is $\dot{B}_t(1)$ 
for a fixed $q$ and $c$. By scanning through possible values of $\dot{B}_t(1)$, 
we numerically find the value $\dot{B}_{tc}(1)$ satisfying the asymptotic boundary 
condition~(\ref{boundary-con}) for each $q$ and $c$~\footnote{We actually solve the two equations 
from $u=1-\epsilon$ located just outside the horizon and use $Y(1-\epsilon)$, 
$\dot{Y}(1-\epsilon)$, $B_{t}(1-\epsilon)$, and $\dot{B}_t(1-\epsilon)$ as initial data. 
Here, $\epsilon$ is small and positive. These initial values are obtained by 
expanding $Y(u)$ and $B_t(u)$ around $u=1$. The numerical results do not 
depend on $\epsilon$. }. 

For later convenience, we shall introduce a parameter $\xi$ defined by 
\begin{align}
\label{def-xi}
c=\xi+\xi^2+\xi^3. 
\end{align}
Then, the Cauchy horizon~(inner horizon)~radius $u_i$ is represented 
by $\xi~(<1)$ as $u_i=1/\xi$. 
In Figs.~1, 2, we give the numerical results for the short wavelength~($q=3$) 
and long wavelength~($q=0.5$) cases at $\xi=0.5$, respectively.   
 
\begin{figure}
 \begin{center}
  \includegraphics[width=7truecm,clip]{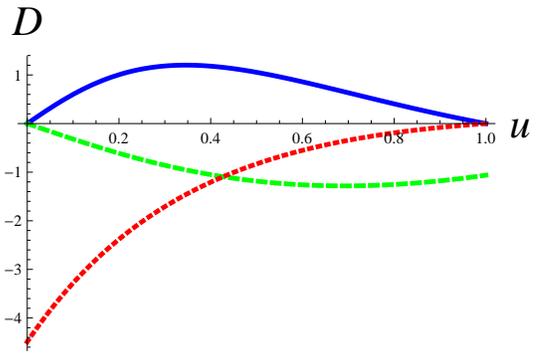}
  \caption{(color online) $D=b(u)$~(solid curve), $F(u)$~(dashed curve), 
and $B_t(u)\times 10^{-1}$~(dotted curve) are shown, 
respectively for $\xi=0.5$, $q=3$.} 
 \end{center}
\end{figure}

\begin{figure}
 \begin{center}
  \includegraphics[width=7truecm,clip]{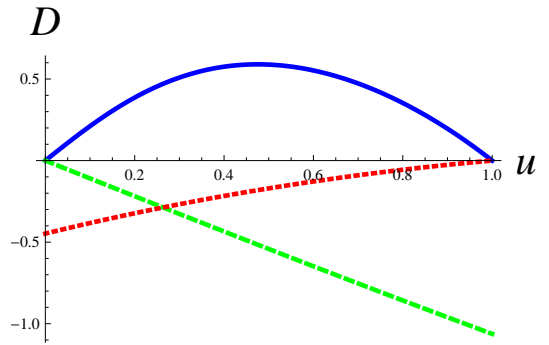}
  \caption{(color online) 
$D=b(u)$~(solid curve), $F(u)$~(dashed curve), and 
$B_t(u)\times 10^{-1}$~(dotted curve) are shown, respectively for $\xi=q=0.5$.} 
 \end{center}
\end{figure}

We also solve Eqs.~(\ref{two-coupled-eq_non-extreme})
from the event horizon towards the Cauchy horizon $u_i=1/\xi$
of the black hole.
We numerically find that, for any wave number $q$,
$F(u)$ blows up towards the Cauchy horizon,
and that the gauge invariant part $\delta \calK_q(u)$ of the perturbation
of the Kretschmann scalar curvature invariant also does.
%
Figs.~3 and 4 show 
the numerical results for the same parameters as in Figs.~1, 2.

In the long wavelength limit, $q\to 0$,
$\delta\calK_q(u)$ can be expanded as
$\delta\calK_q(u) = I_0(u) + q^2\, I_1(u) + O(q^4)$. 
Substituting the solution constructed in Sec.~III into
$\delta\calK_q(u)$, 
we find that $I_0(u)$ is finite at $u_i=1/\xi$, but $I_1(u)$ blows
up when $u\to 1/\xi$ as 
\begin{align}
I_1(u)\sim-\frac{24(1+\xi)^2(1+\xi^2)(5-\xi^3(8-3\xi))}
{L^4\xi^6(1+\xi(2+3 \xi))^2} \ln(1-\xi u). 
\end{align}
%
These numerical and analytical results
imply that the inhomogeneous charged black hole geometries
terminate at singularity instead of the appearance of the Cauchy horizon.

Inside the event horizon, the timelike killing vector becomes spacelike.
If one compactifies the three spacelike directions, $t$, $x$, and $y$,
the spacetime becomes $R^1\times T^3$ Gowdy universe~\cite{Gowdy}
with two commuting spacelike killing vector fields, $\p_t$ and $\p_y$.
As shown numerically and analytically
in the vacuum case~\cite{Moncrief,Berger},
such a universe generically terminates at spacelike curvature singularity,
supporting the strong cosmic censorship conjecture.
These results would not change even if electromagnetic field exists
because the dominant energy condition is still satisfied.
So, our analysis is consistent with the results~\cite{Moncrief,Berger}
and we conjecture that spacelike singularity generically appears
instead of the appearance of the Cauchy horizon
in the inhomogeneous charged black hole solutions.

\begin{figure}
 \begin{center}
  \includegraphics[width=7truecm,clip]{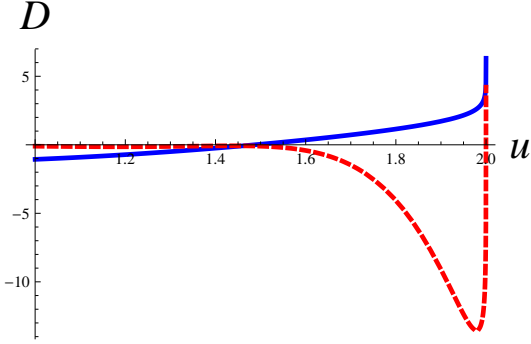}
  \caption{(color online) 
$D=F(u)$~(solid curve) and $L^4 {\mathcal K}_1(u)\times 10^{-3}$~(dashed curve) are shown, 
respectively for $\xi=0.5$, $q=3$.} 
 \end{center}
\end{figure}

\begin{figure}
 \begin{center}
  \includegraphics[width=7truecm,clip]{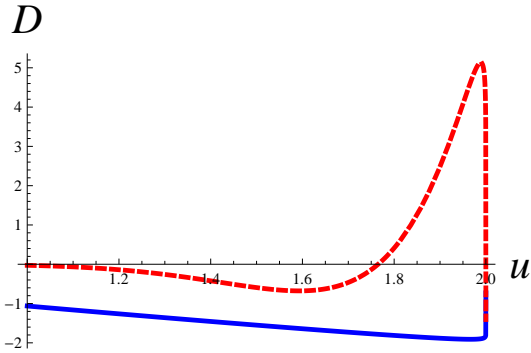}
  \caption{(color online) 
$D=F(u)$~(solid curve) and $L^4 {\mathcal K}_1(u)\times 10^{-3}$~(dashed curve) are 
shown, respectively for $\xi=q=0.5$.} 
 \end{center}
\end{figure}

\section{extremal solutions}
In this section, we investigate extremal inhomogeneous solutions of the two 
coupled equations~(\ref{two-coupled-eq_non-extreme}). Introducing new variable 
$\Gamma(u)$ as $\Gamma=B_t(u)/(1-u)$ and substituting $c=3$, 
Eqs.~(\ref{two-coupled-eq_non-extreme}) are rewritten as 
\begin{subequations}
\begin{align}
& (1-u)^2(1+2u+3u^2)(1+u+u^2+3u^3)\ddot{Y}(u) \nonumber \\
&-6(1-u)u^2(6+5u+4u^2+3u^3)\dot{Y}(u) \nonumber \\
&-q^2(1+u+u^2+3u^3)Y(u)-12u\Gamma(u)=0, 
\label{eq-Yex} \\
& (1-u)^2(1+u+u^2+3u^3)^2(1+2u+3u^2)\ddot{\Gamma}(u) \nonumber \\
& -2(1-u)(1+u+u^2)(1+u+u^2+3u^3)\times \nonumber \\
&\,\,\,(1+2u+3u^2)\dot{\Gamma}(u) \nonumber \\
& -[q^2(1+u+u^2+3u^3)^2 \nonumber \\
&\,\,\,+6u^2(3+u+u^2+u^3)(1+2u+3u^2)]\Gamma(u) \nonumber \\
& +3(1-u)^3(1+2u+3u^2)^3\dot{Y}(u) \nonumber \\
&-3q^2u(1+u+u^2+3u^3)(1+2u+3u^2)Y(u)=0. 
\label{eq-G}
\end{align} 
\label{two-coupled-eq-extreme}
\end{subequations}

These equations~(\ref{two-coupled-eq-extreme}) can be transformed into four coupled 
first-order differential equations by introducing two variables, $P$ and $Q$ as 
\begin{align}
P(u):=(1-u)\dot{Y}(u), \qquad Q(u):=(1-u)\dot{\Gamma}(u). 
\end{align} 
Near the event horizon, $u=1$,  the four coupled 
first-order differential equations are 
represented as a regular matrix form:  
\begin{align}
\label{matrix-form}
(1-u)\dot{{\bm X}}
=M{\bm X}\simeq 
\left(
\begin{array}{cccc}
0 & 0 & 1 & 0 \\
0 & 0 & 0 & 1 \\
q^2/6 & 1/3 & 2 & 0 \\
q^2/2 & 1+q^2/6 & 0 & 0 \\
\end{array}
\right)
{\bm X}, 
\end{align}
where ${\bm X}=(Y,\,\Gamma,\,P,\,Q)$. 

To find the indices characterizing the behavior of the solution near $u=1$, 
we substitute the ansatz ${\bm X}_\lambda(u)=(1-u)^\lambda {\bm H}(u;\lambda)$ 
into Eq.~(\ref{matrix-form}). Thus, finding the indices $\lambda$ is equivalent 
to finding the eigenvalues $\lambda$ for the matrix $M$, i.~e.~, 
$|M+\lambda I|=0$. The four eigenvalues and the corresponding eigenvectors 
${\bm H}(1;\lambda)$ are given by  
\begin{subequations}
\begin{align}
& \lambda_{nn\pm}=-\frac{1}{6}\left[3+ \sqrt{3}\sqrt{15+2q^2
\pm 4\sqrt{3}\sqrt{3+q^2}}\, \right], \nonumber \\
& \lambda_{n\pm}=-\frac{1}{6}\left[3- \sqrt{3}\sqrt{15+2q^2
\pm 4\sqrt{3}\sqrt{3+q^2}}\, \right], 
\label{eigenvalues} \\
& \lim_{u\to 1}{\bm H}(u;\lambda)=(1, G(\lambda) , -\lambda, -\lambda G(\lambda)), 
\label{eigenvectors}
\end{align}
\end{subequations}
where 
\begin{align}
  & G(\lambda)
  := 3\, \lambda^2 + 6\, \lambda - \frac{q^2}{2}~.
\label{def-G}
\end{align}
%

We can obtain all the variables $Y$, $B_t$, and $F$ from each solution 
${\bm X}_\lambda$ and Eq.~(\ref{eq-constraint1}). It is easily checked that the 
Hamiltonian constraint equation~(\ref{Hamiltonian-constraint}) is automatically 
satisfied for each mode ${\bm H}(u;\lambda)$. Thus, 
the solutions of Eqs.~(\ref{pert-field1-eq}) are constructed from the four 
independent mode solutions, \{${\bm H}(u;\lambda)$\}. 

Since $\lambda_{nn\pm}<0$ for any $q~(\neq 0)$, the mode solutions ${\bm H}(u;\lambda_{nn\pm})$ 
are ``non-normalizable", i.~e.~, the variables $Y$ and $F$ diverge at the horizon. 
Substituting the eigenvectors~(\ref{eigenvectors}) into Eq.~(\ref{curvature-behavior}), 
we can show that the scalar curvature invariant (\ref{curvature-behavior}) 
blows up at the horizon for each ``non-normalizable" mode solution. So, we must 
abandon the ``non-normalizable" solutions by the regularity 
condition~(\ref{horizon-regularity}). 
Since $\lambda_{n\pm}>0$ for any $q~(\neq 0)$, the other two mode solutions 
${\bm H}(u;\lambda_{n\pm})$ are ``normalizable" and Eq.~(\ref{curvature-behavior}) 
is finite at the horizon. 
Then, the ``normalizable" mode solutions ${\bm H}(u;\lambda_{n\pm})$ are permitted 
as a regular solution at the horizon. 

To construct an extremal inhomogeneous black hole solution in an asymptotically 
AdS spacetime, we must impose the asymptotic boundary condition~(\ref{boundary-con}). 
Let us consider the one parameter family of solutions ${\bm H}(u;p)$ of the 
equations~(\ref{two-coupled-eq-extreme}) by superposing the two 
independent mode solutions, ${\bm H}(u;\lambda_{n\pm})$ as  
${\bm H}(u;p)={\bm H}(u;\lambda_{n+})+p{\bm H}(u;\lambda_{n-})$ for a fixed $q$.    
By scanning all possible values of $p$, we find the value $p_c$ satisfying the 
asymptotic condition~(\ref{boundary-con}). We numerically find the solutions ${\bm H}(u;p_c)$ 
for various values of $q$. 
Two typical cases are shown in Fig.~5~($q=3$) and Fig.~6~($q=8$). The former case 
represents the solution where the derivatives of the metric functions diverge at 
the horizon, while the latter case represents a smooth solution at the horizon.    
\begin{figure}
 \begin{center}
  \includegraphics[width=7truecm,clip]{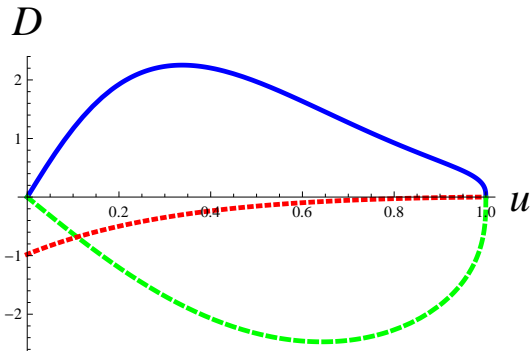}
  \caption{(color online) 
$D=b(u)$~(solid curve), $F(u)$~(dashed curve), and $B_t(u)/10^2$~(dotted curve) 
are shown, respectively for $q=3$.} 
 \end{center}
\end{figure}
\begin{figure}
 \begin{center}
  \includegraphics[width=7truecm,clip]{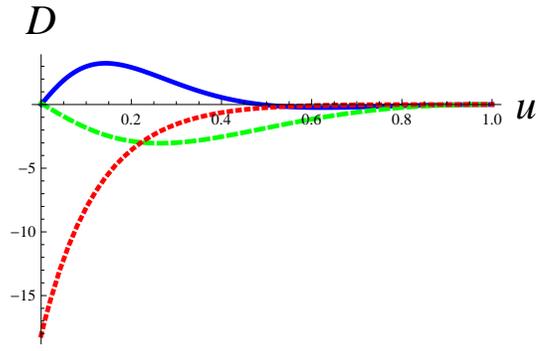}
  \caption{(color online) 
$D=b(u)$~(solid curve), $F(u)$~(dashed curve), and $B_t(u)/10^2$~(dotted curve) 
are shown, respectively for $q=8$.} 
 \end{center}
\end{figure}
As shown below, the singular behavior in the former case induces a curvature 
singularity at the horizon even though Eq.~(\ref{cond-curvature}) is satisfied. 
For each extremal black hole solutions with wave number $q$, ${\bm H}(u;{p_c})$ 
generically includes the mode solution ${\bm H}(u, \lambda_{n-})$, 
as $p_c$ is not generically zero. Since $\lambda_{n-}<\lambda_{n+}$, 
the singular behavior must be encoded in the the mode solution ${\bm H}(u, \lambda_{n-})$. 

We calculate the curvature component in the frame parallelly propagated  
along a freely falling observer into the event horizon, for the mode solution 
${\bm H}(u, \lambda_{n-})$. 
If we take the real part of the metric, the tangential vector ${\bm V}_0$ of 
the timelike 
geodesic of the freely falling observer within $x=y=0$ is given by  
\begin{align}
\label{tangentvector}
& V_0^t=\dot{t}=\frac{E u^2}{L^2 g(u)}(1+\epsilon b(u)), 
\nonumber \\
& V_0^u=\dot{u}=\sqrt{\frac{E^2u^4}{L^4}-\frac{u^2g(u)}{L^2}(1-\epsilon b(u))}, 
\nonumber \\
& V_0^x=V_0^y=0, 
\end{align}
where $E$ is a positive constant and a dot means the derivative with respect to 
the proper time $\tau$ of the observer. In terms of the orthogonal frame 
${\bm E}_n$~(up to $O(\epsilon)$)
\begin{align}
\label{orthogonal-frame}
& {\bm E}_0=\frac{u}{L\sqrt{g(u)}}\left(1+\frac{\epsilon}{2}b(u)\cos qx \right)\p_t, 
\nonumber \\
& {\bm E}_1=\frac{u\sqrt{g(u)}}{L}\left(1-\frac{\epsilon}{2}b(u)\cos qx \right) \p_u, 
\nonumber \\
& {\bm E}_2=\frac{u}{L}\left(1-\epsilon F(u)\cos qx \right)\p_x, 
\nonumber \\
& {\bm E}_3=\frac{u}{L}\left(1-\epsilon F(u)\cos qx \right) \p_y, 
\end{align}
a boost parameter $\alpha$ is defined by 
\begin{align}
{\bm V}_0=\cosh\alpha\, {\bm E}_0+\sinh\alpha\, {\bm E}_1. 
\end{align}

Some curvature components relevant below are calculated as 
\begin{align}
\label{riemann}
& R_{txtx}=\frac{L^2g(u)}{2u^4}(2g(u)-u\dot{g}(u)) \nonumber \\
&+\epsilon \cos qx\frac{L^2g(u)}{2u^4}
[Y(u)(q^2u^2-4g(u)+2u\dot{g}(u)) \nonumber \\
&+2F(u)(q^2u^2-2g(u)+u\dot{g}(u)) \nonumber \\
&+u^2\dot{g}(u)\dot{F}(u)+ug(u)\dot{Y}(u)]+O(\epsilon^2), \nonumber \\
& R_{uxux}=-\frac{L^2}{2u^4g(u)}(2g(u)-u\dot{g}(u)) \nonumber \\
&+\frac{\epsilon L^2 \cos qx}{2u^4g(u)}
[2F(u)(q^2u^2-2g(u)+u\dot{g}(u))+u^2q^2Y(u) \nonumber \\
&-u^2\dot{g}(u)\dot{F}(u)-ug(u)(\dot{Y}(u)+2u\ddot{F}(u)]+O(\epsilon^2) \nonumber \\
&=-\frac{L^2}{2u^4g(u)}(2g(u)-u\dot{g}(u)) \nonumber \\
&-\frac{\epsilon L^2\cos qx}{2u^4g(u)}[F(u)(4g(u)-2u\dot{g}(u)) \nonumber \\
&+u^2\dot{g}(u)\dot{F}(u)+ug(u)\dot{Y}(u)]+O(\epsilon^2), \nonumber \\
& R_{txux}=0, 
\end{align}  
where we used Eq.~(\ref{eq-F1}) in the second equality of $R_{uxux}$. 
Thus, near the horizon, the curvature component 
$R_{\mu\nu\alpha\beta}V^\mu_0 E^\nu_2 V^\alpha_0 E^\beta_2$~\footnote{Along the 
timelike geodesic with tangent vector ${\bm V}_0$, ${\bm E}_2$ is a basis in the parallelly 
propagated frame.} in the parallelly propagated frame is calculated as   
\begin{align}
& R_{\mu\nu\alpha\beta}V^\mu_0 E^\nu_2 V^\alpha_0 E^\beta_2=\cosh^2\alpha 
\,R_{\mu\nu\alpha\beta} E^\mu_0 E^\nu_2 E^\alpha_0  E^\beta_2 \nonumber \\
&+\sinh^2\alpha\, R_{\mu\nu\alpha\beta}
E^\mu_1 E^\nu_2 E^\alpha_1 E^\beta_2 \nonumber \\
&\simeq 
\cosh^2 \alpha \left(\frac{\epsilon q^2 b(u)}{L^2}+O((1-u)^{\lambda_{n-}+1})\right) 
\nonumber \\
&\sim \frac{\epsilon E^2q^2}{L^4}(1-u)^{\lambda_{n-}-2}. 
\end{align} 
Here, we used $\cosh^2\alpha\simeq E^2/L^2g(u)$ near the horizon. 
Note that 
this divergence is a peculiar phenomenon for the inhomogeneous solution, as 
it originates in the curvature component $R_{txtx}$ at $O(\epsilon)$.  
Thus, the inhomogeneous extremal solution generically includes 
{\it p.~p. curvature singularity}~\cite{Hawking} at the event horizon 
when 
\begin{align}
\label{weak-con}
\lambda_{n-}<2 \,\,\, \mbox{i.~e.,} \,\,\, |q|<\sqrt{36+12\sqrt{3}}. 
\end{align}
This implies that we cannot smoothly extend the geometry inside 
the event horizon even though the Kretschmann scalar curvature invariant 
remains small. 

As shown below, we can see that this curvature singularity is a strong 
curvature singularity when 
\begin{align}
\label{con-wave-number}
\lambda_{n-}<1 \,\,\, \mbox{i.~e.,} \,\,\, |q|< 2\sqrt{6}
\end{align} 
is satisfied.  
Let us consider a timelike geodesic congruence with the tangent vector field ${\bm V}$ 
including the timelike geodesic with the tangent vector ${\bm V}_0$. 
The shear tensor obeys the evolution equation along the timelike congruence as 
\begin{align}
& \frac{d\sigma_{22}}
{d\tau}\sim C_{\mu\nu\alpha\beta}
V^\mu_0 E^\nu_2 V^\alpha_0 E^\beta_2 \sim (1-u)^{\lambda_{n-}-2}.  
\end{align}
So, the shear tensor in the parallelly propagated frame
diverges for {\it any} timelike geodesic congruence 
when (\ref{con-wave-number}) is satisfied. 
In this sense, the p.~p. curvature singularity is a strong curvature singularity.

The p.~p. curvature singularity has been observed in a class of extremal
black holes in string theory~\cite{horowitzross1997, horowitzross1998}.
This is caused by the divergence of the stress-energy tensor of the dilaton field at the 
horizon, while in our case, the matter field associated with the gauge 
field $\delta F_{\mu\nu}$ does not diverge there.   
Therefore, the origin of the p.~p. curvature singularity is different between the 
inhomogeneous extremal black hole and the extremal black holes in string theory.

\section{Conclusion and discussions}
We have investigated four-dimensional inhomogeneous charged black hole
solutions in the anti-de Sitter backgrounds in the Einstein-Maxwell system.
In the framework of linear perturbations, we have constructed the solutions 
where the inhomogeneity is induced by a spatially inhomogeneous 
chemical potential with wave number $q$. 
For long wavelength limit, $q\to 0$, an analytic solution is constructed,
up to $O(q^2)$. 
By superposing the solutions with different wave numbers,
we can obtain an inhomogeneous charged black hole solution
for an arbitrary configuration of the chemical potential. 

At the extremal case, p.~p.~curvature singularity generically appears
at the event horizon for the long wavelength perturbations
even though the Kretschmann scalar curvature invariant 
$R_{\mu\nu\alpha\beta}R^{\mu\nu\alpha\beta}$ remains small. 
This implies that any freely-falling observer into
the inhomogeneous black hole feels infinite tidal force at the event horizon.
As the shear of any timelike geodesic congruence of the freely-falling
observer diverges infinitely~(Sec.~IV), 
the p.~p.~curvature singularity is a strong curvature singularity
and the geometry cannot be smoothly extended into the inside of the black hole. 
Quite recently, it has been shown that the Lifshitz spacetime possessing the same 
p.~p.~curvature singularity is unstable against quantum corrections 
in string theory~\cite{horowitzway}. This suggests that the extremal black holes 
are also generically unstable against the quantum corrections, even though 
the extremal Reissner-Nordstr\"{o}m AdS solution is stable.  

\begin{figure}[htbp]
 \begin{minipage}{0.43\hsize}
  \begin{center}
   \includegraphics[width=27mm]{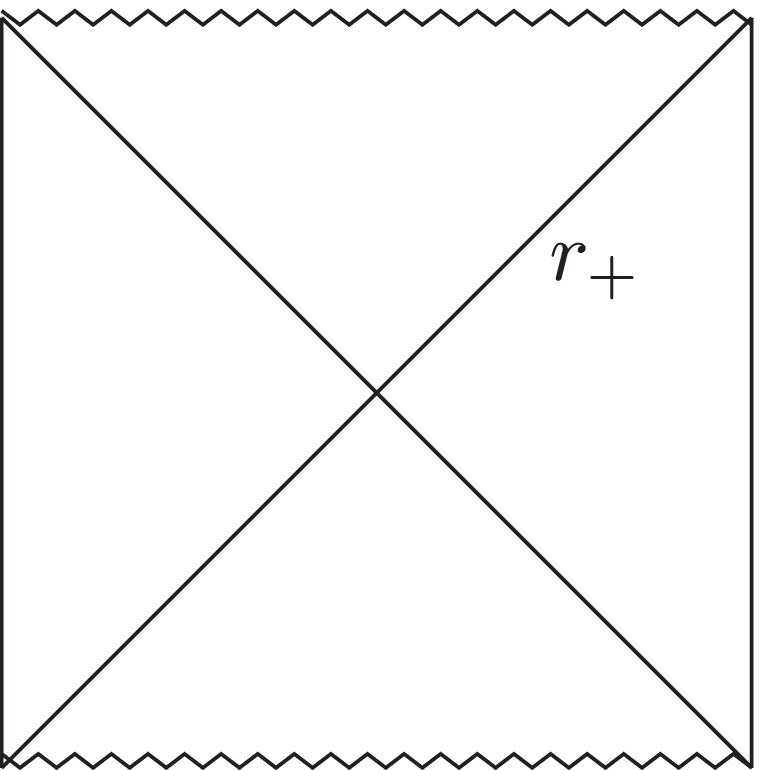}
  \end{center}
  \caption{\small{The causal structure of Schwarzschild-AdS spacetime. $r_+$ is 
the event horizon.}}
  \label{Sch-AdS}
 \end{minipage}
 \begin{minipage}{0.43\hsize}
 \begin{center}
  \includegraphics[width=27mm]{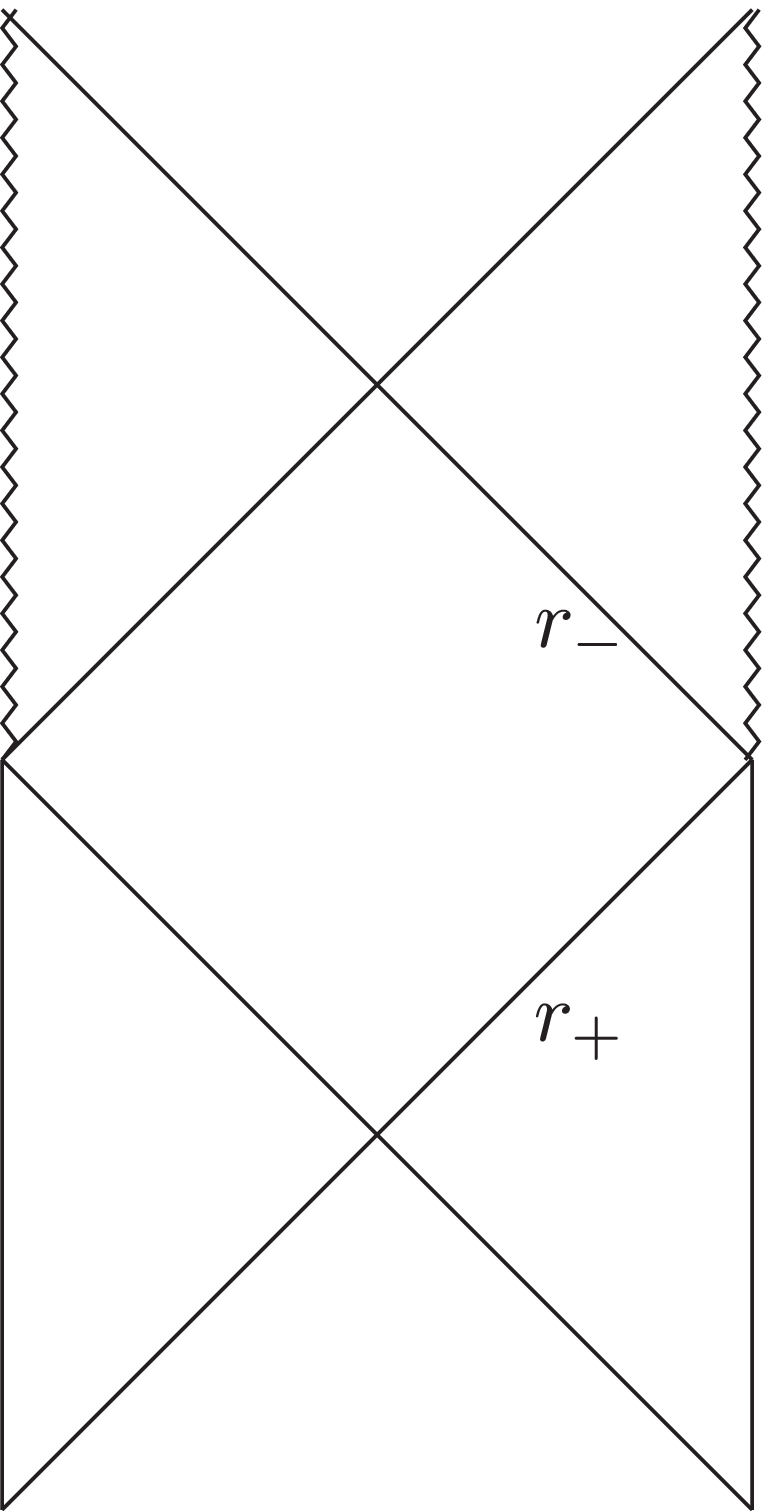}
 \end{center}
  \caption{\small
  {The causal structure of Reissner-Nordstr\"{o}m-AdS spacetime. $r_-$ is the 
Cauchy horizon.}}
  \label{RN-AdS}
 \end{minipage}
\end{figure}

The generic appearance of the p.~p.~curvature singularity
would be associated with the inner causal structure
of the non-extremal solutions.
As shown in Sec.~III, the perturbation with any wave length breaks down
at the Cauchy horizon and the scalar curvature grows
towards the Cauchy horizon. As discussed in Sec.~III,
this curvature growth suggests that the inner causal structure
of the inhomogeneous black hole solution is similar
to the one of the Schwarzschild-AdS spacetime~(Fig.~\ref{Sch-AdS})
rather than the unperturbed Reissner-Nordstr\"{o}m-AdS spacetime~
(Fig.~\ref{RN-AdS}).
Thus, the curvature at the event horizon grows
as the Cauchy horizon approaches the event horizon. 
However, it is not still clear why short wavelength 
perturbation~(large $q$ case) does not yield
the p.~p.~curvature singularity.
It would be interesting to explore whether or not
the perturbation of Reissner-Nordstr\"{o}m-AdS spacetime
with spherical or hyperbolic horizon possesses
the p.~p.~curvature singularity at the extremal limit. 

Finally we refer to the AdS/CFT duality.
According to the AdS/CFT duality,
a charged black hole solution is
dual to the boundary field theory at a finite temperature
with a finite chemical potential.
An inhomogeneous chemical potential induces
an \lq\lq electric\rq\rq\, force in the charged matter
of the boundary field theory.
In our equilibrium configurations,
we can show that
this \lq\lq electric\rq\rq\, force balances
with the pressure gradient,
\begin{align}
   \partial_x \Exp{T^{xx}}
  &= \Exp{J^t}\, \calE_x~,
\label{eq:div-calT^x}
\end{align}
where $\Exp{T^{ab}}$ and $\Exp{J^a}$~~($a, b = t, x, y$)
are the expectation values of the energy-momentum tensor
and the conserved current of the dual field theory,
and $\calE_x$ is the \lq\lq electric\rq\rq\, field
$\calE_x := \lim_{u \to 0} F_{tx}$ on the boundary.
The equation (\ref{eq:div-calT^x}) means that
the pressure gradient is balanced with the \lq\lq electric\rq\rq\, force
(see appendix \ref{sec:force_balance} for details).
It is one of the reasons why the Einstein-Maxwell system permits
such inhomogeneous black hole solutions
under the asymptotically AdS boundary condition,
while the Einstein vacuum system does not permit them%
\footnote{If we relax the asymptotically AdS boundary condition,
there exists the static inhomogeneous solutions
even in the Einstein vacuum system.
In that cases, the pressure gradient balances with
the external gravitational force on the boundary.
}
\cite{Anderson,yoshino}. 

According to the AdS/CFT duality, our black hole solution is dual
to the strongly coupled gauge theory under the periodic
chemical potential in a flat $2+1$-dimensional spacetime.
In condensed matter physics, such a periodic structure
is a key ingredient to understand the energy gap or the band structure.
So, it would be interesting to investigate
the mechanism of energy dissipation
in the periodic charged black hole background.

\begin{acknowledgments} 
This research was supported in part by the Grant-in-Aid for Scientific
Research~No.~23740200
and No.~23540326 
from the Ministry of Education, Culture,
Sports, Science and Technology, Japan.
\end{acknowledgments}

\appendix
\section{force balance equation}
\label{sec:force_balance}
It is convenient to derive various expectation values
of a dual field theory in the framework of ADM-like decomposition
in which a spacetime $\calM$ is foliated by timelike hypersurfaces 
homeomorphic to the AdS boundary $\partial \calM$:
\begin{align}
   ds_4^2
  &= N^2\, du^2 + \gamma_{ab} \left( dx^a + N^a\, du \right)
    \left( dx^b + N^b\, du \right)
\label{eq:ADM_decomposition} \\
  &\xrightarrow{u \to 0}\,
  \frac{L^2}{u^2}\, \left[\, du^2 + \eta_{ab}\, dx^a\, dx^b
  + O\left( u^3 \right)\, \right]~, 
\nonumber
\end{align}
Here, the AdS boundary is located at $u=0$ and
$x^a$ are the coordinates on the $u=\mbox{const.}$ timelike hypersurface
$\partial \calM_u$. The induced metric $\gamma_{ab}$ on $\partial \calM_u$ 
is given by $\gamma_{\mu\nu} := g_{\mu\nu} - n_\mu n_\nu$,
where $n^\mu$ is the unit normal to $\partial \calM_u$
and is chosen to be ``outward pointing''. 

Using the extrinsic curvature
$K^{\mu\nu} := \gamma^{\mu\lambda}\, \nabla_\lambda n^\nu$
of $\partial \calM_u$ and the holographic counter-term $S_{\text{ct}}$, 
the expectation values $\Exp{T^{ab}}$ and $\Exp{J^a}$ are expressed by
$\Exp{T^{ab}} = \lim_{u \to 0} \calT^{ab}$
and $\Exp{J^a} = \lim_{u \to 0} \calJ^a$, respectively, where
\begin{align}
   \calT^{ab}
  &= 2\, \left( \frac{L}{u} \right)^{5}
  \left( \gamma^{ab} K - K^{ab}
    + \frac{1}{\sqrt{-\gamma}}\,
      \frac{\delta S_{\text{ct}}}{\delta \gamma_{ab}} \right)~,
\label{eq:def-calT^ab} \\
   \calJ^a
  &= \left( \frac{L}{u} \right)^{3} F^{a\mu}\, n_\mu~.
\label{eq:def-calJ^a}
\end{align}
$S_{\text{ct}}$ must be chosen
to cancel divergences that arise as $\partial \calM_u$
approaches the AdS boundary $\partial \calM$.
In the following argument, we need the properties of $S_{\text{ct}}$
which is invariant under the diffeomorphism on $\partial \calM_u$. 
We may, however, refer to Ref.~\cite{Batrachenko:2004fd} 
for the concrete expression of $S_{\text{ct}}$. 

The conservation laws of $\Exp{T^{ab}}$ and $\Exp{J^a}$ are encoded
by the constraints in the bulk side,
\begin{align}
   0
  &= - 2\, D_b \left( \gamma^{ab} K - K^{ab} \right)
  + \left( F^{b\mu}\, n_\mu \right) F^a{}_{b}
  ~,
\label{eq:def-hcH_a} \\
   0
  &= - D_a \left( F^{a\mu}\, n_\mu \right)~,
\label{eq:def-hPhi_I}
\end{align}
where $D_a$ is the covariant derivative with respect to
the induced metric $\gamma_{ab}$ on $\partial \calM_u$.

Since the Christoffel symbol $\Gamma^a_{bc}[\, \gamma\, ]$
with respect to $\gamma_{ab}$ behaves as 
$\Gamma^a_{bc}[\, \gamma\, ] = O(u^3)$,
the current conservation law holds as
\begin{align*}
  & \partial_a \Exp{J^a}
  = \lim_{u \to 0}\, D_a \calJ^a
  = \lim_{u \to 0}\, \left( \frac{L}{u} \right)^{3}\, D_a
  \left( F^{a\mu}\, n_\mu \right)
  = 0~.
\end{align*}
Similarly, we obtain
\begin{align}
   \partial_b \Exp{T^{ab}}
  &= \lim_{u \to 0}\, D_b \calT^{ab}
  = 2\, \lim_{u \to 0}\, \left( \frac{L}{u} \right)^{5} D_b
  \left( \gamma^{ab} K - K^{ab} \right)
\nonumber \\
  &= \lim_{u \to 0}\, \left( \frac{L}{u} \right)^{5}
  \left( F^{b\mu}\, n_\mu \right) F^a{}_{b}
\nonumber \\
  &= \Exp{J^b}\, \left( \lim_{u \to 0}\, \eta^{ac}\, F_{cb} \right)
  ~,
\label{eq:div-calT}
\end{align}
where we make use of $D_b \left[\, \left( 1/\sqrt{-\gamma} \right)\,
\left( \delta S_{\text{ct}}/\delta \gamma_{ab} \right)\, \right] = 0$
because $S_{\text{ct}}$ is invariant under the diffeomorphism
on $\partial \calM_u$, $\gamma_{ab}
\to \gamma_{ab} - 2\, D_{(a} \xi_{b)}$.
Thus, we get the force-balance equation (\ref{eq:div-calT^x}) 
for the static inhomogeneous configurations in the $x$-direction. 
\section{Solution at $O(q^2)$}
In the long wavelength limit,
an explicit inhomogeneous charged black hole solution is given,
up to $O(q^2)$, as 
\begin{align}
& 
F_1(u)
=\frac{\arctan\left(\frac{\xi +1}{\sqrt{3+2\xi+3\xi^2}}\right)
-\arctan\left(\frac{1+\xi+2u(1+\xi+\xi^2)}{\sqrt{3+2\xi+3\xi^2}}\right)}
{(1+2\xi+3\xi^2)^3(3+2\xi+3\xi^2)^{3/2}}\times \nonumber \\
& [6+50\xi+2\xi^2(27 \xi ^5+63 \xi ^4+111 \xi ^3+129 \xi ^2+109 \xi +61) \nonumber \\
& +3u(1+\xi)(1+\xi^2)(9 \xi ^5+12 \xi ^4+16 \xi ^3+10 \xi ^2+7 \xi +2)] \nonumber \\
& +\{\ln(1+(1+\xi)u+(1+\xi+\xi^2)u^2)-2\ln(1-\xi u)\}\times \nonumber \\
&\frac{(1+\xi)[2+2\xi(1+\xi+3\xi^2)-3\xi(1+\xi)(1+\xi^2)u]}{2(1+2\xi+3\xi^2)^3} \nonumber \\
&+\frac{8\xi(1+\xi+\xi^2)u}{(1+2\xi+3\xi^2)^2(3+2\xi+3\xi^2)}.  
\end{align}
\begin{align}
& B_{t1}(u)=\frac{1}{2(1+2\xi+3\xi^2)^3(3+2\xi+3\xi^2)}\times \nonumber \\
&\Biggl[4(1-u)(1+2\xi+3\xi^2)\times \nonumber \\
&(3+20\xi(1+\xi+\xi^2)+\xi^2(1+\xi+\xi^2)^2(25-8u))
\nonumber \\
&-3(1-\xi^2)^3(1+\xi^2)(3+2\xi+3\xi^2)\ln\left\{\frac{(1-\xi)^2}{3+2\xi+\xi^2}\right\}\Biggr]
\nonumber \\
& +\frac{3(2\ln(1-\xi u)-\ln\{1+(1+\xi)u+(1+\xi+\xi^2)u^2\}}
{2(1+2\xi+3\xi^2)^3}\times \nonumber \\
&(1+\xi)^2(1+\xi^2)(1-\xi u)\{1+\xi+\xi^2+3\xi^3-2\xi(1+\xi+\xi^2)u\} \nonumber \\
&+3\frac{(1+\xi)(1+\xi^2)
\arctan\left(\frac{1+\xi+2u(1+\xi+\xi^2)}{\sqrt{3+2\xi+3\xi^2}}\right)}
{(1+2\xi+3\xi^2)^3(3+2\xi+3\xi^2)^{3/2}}\times \nonumber \\
& [3+u\{3(1+\xi)(1+\xi^2)-2u\xi(1+\xi+\xi^2)\} \nonumber \\
& \,\,\,\,\times(2+7\xi+10\xi^2+16\xi^3+12\xi^4+9\xi^5) \nonumber \\
& \,\,\,+\xi(25+61\xi+109\xi^2+129\xi^3+111\xi^4+63\xi^5+27\xi^6)] \nonumber \\
& +3\frac{(1+\xi)(1+\xi^2)
\arctan\left(\frac{1+\xi}{\sqrt{3+2\xi+3\xi^2}}\right)}
{(1+2\xi+3\xi^2)^3(3+2\xi+3\xi^2)^{3/2}}\times  \nonumber \\
& \{3+\xi+\xi^2+\xi^3-3u(1+\xi)(1+\xi^2)+2u^2\xi(1+\xi+\xi^2)\} \nonumber \\
& \times (2+7\xi+10\xi^2+16\xi^3+12\xi^4+9\xi^5) \nonumber \\
& -3\frac{(1+\xi)(1+\xi^2)
\arctan\left(\frac{3+3\xi+2\xi^2}{\sqrt{3+2\xi+3\xi^2}}\right)}
{(1+2\xi+3\xi^2)^3(3+2\xi+3\xi^2)^{3/2}}\times  \nonumber \\
& (9+48\xi+100\xi^2+176\xi^3+198\xi^4+176\xi^5 \nonumber \\
&\,\,\,\,\,+100\xi^6+48\xi^7+9\xi^8). \nonumber \\
\end{align}
\clearpage
\begin{align}
& b_1(u):=Y_1(u) + 2 
F_1(u)
\nonumber \\
&=-\frac{6u(1+\xi)^2(1-\xi u)(1+\xi^2)\{\xi+2\xi^2 u-(1+\xi+\xi^2)u^2\}}
{(1-u)(1+2\xi+3\xi^2)^3(1+(1+\xi)u+(1+\xi+\xi^2)u^2)}\ln(1-\xi u) \nonumber \\
&-\frac{6u^3(1-\xi^2)^3(1+\xi^2)\ln(1-\xi)}
{(1-u)(1-\xi u)(1+2\xi+3\xi^2)^3(1+(1+\xi)u+(1+\xi+\xi^2)u^2)} \nonumber \\
& -\frac{6(1+\xi)(1+\xi^2)\arctan\left[\frac{1+\xi}{\sqrt{3+2\xi+3\xi^2}}  \right]}
{(1-\xi u)(1+2\xi+3\xi^2)^3(3+2\xi+3\xi^2)^{3/2}\{1+(1+\xi)u+(1+\xi+\xi^2)u^2\}}
\nonumber \\
& \times u(1-u)(1+2u-\xi(1+\xi+\xi^2)u^2)
(2+7\xi+10\xi^2+16\xi^3+12\xi^4+9\xi^5) \nonumber \\
&-\frac{6(1+\xi)(1+\xi^2)u^3\arctan\left[\frac{3+3\xi+2\xi^2}{\sqrt{3+2\xi+3\xi^2}}  \right]}
{(1-u)(1-\xi u)(1+2\xi+3\xi^2)^3(3+2\xi+3\xi^2)^{3/2}\{1+(1+\xi)u+(1+\xi+\xi^2)u^2\}}
\nonumber \\
&\times (9+48\xi+100\xi^2+176\xi^3+198\xi^4+176\xi^5+100\xi^6+48\xi^7+9\xi^8)
\nonumber \\
& +\frac{6(1+\xi)(1+\xi^2)u\arctan\left[\frac{1+\xi+2u(1+\xi+\xi^2)}
{\sqrt{3+2\xi+3\xi^2}}  \right]}
{(1-u)(1-\xi u)(1+2\xi+3\xi^2)^3(3+2\xi+3\xi^2)^{3/2}\{1+(1+\xi)u+(1+\xi+\xi^2)u^2\}}
\nonumber \\
& \times [(2+7\xi+10\xi^2+16\xi^3+12\xi^4+9\xi^5)
(1+2(1+\xi)(1+\xi^2)u^3-\xi(1+\xi+\xi^2)u^4) 
\nonumber \\
&\,\,\,\,\,+(3+25\xi+61\xi^2+109\xi^3+129\xi^4+111\xi^5+63\xi^6+27\xi^7)u^2]
\nonumber \\
& +\frac{3u^3(1-\xi^2)^3(1+\xi^2)\ln(3+2\xi+\xi^2)}
{(1-u)(1-\xi u)(1+2\xi+3\xi^2)^3\{1+(1+\xi)u+(1+\xi+\xi^2)u^2\}}
\nonumber \\
& +\frac{3(1+\xi)^2(1+\xi^2)\ln\{(1+(1+\xi)u+(1+\xi+\xi^2)u^2\}}
{(1-u)(1+2\xi+3\xi^2)^3\{1+(1+\xi)u+(1+\xi+\xi^2)u^2\}}
\nonumber \\
&\times u(1-\xi u)\{\xi+2\xi^2 u-(1+\xi+\xi^2)u^2)\}
\nonumber \\
& -\frac{2u^2}{(1-\xi u)(1+2\xi+3\xi^2)^2(3+2\xi+3\xi^2)
\{1+(1+\xi)u+(1+\xi+\xi^2)u^2\}}\times \nonumber \\
&[3(1+\xi)(1+\xi^2)\{1+3\xi(1+\xi+\xi^2)\}+8\xi^2(1+\xi+\xi^2)^2u^2 \nonumber \\
&-u\{3+\xi(1+\xi+\xi^2)(24+29\xi(1+\xi+\xi^2))\}]. 
\end{align}




\begin{thebibliography}{99}
\bibitem{Maldacena:1997re}
  J.~M.~Maldacena,
  ``The large N limit of superconformal field theories and supergravity,"
  Adv.\ Theor.\ Math.\ Phys.\  {\bf 2} (1998) 231
  [Int.\ J.\ Theor.\ Phys.\  {\bf 38} (1999) 1113]
  [arXiv:hep-th/9711200].  

\bibitem{Hartnoll:2008kx}
  S.~A.~Hartnoll, C.~P.~Herzog and G.~T.~Horowitz,
  ``Holographic Superconductors,''
  JHEP {\bf 0812}, 015 (2008)
  [arXiv:0810.1563 [hep-th]].

\bibitem{Lemos1}
J. P. S. Lemos,
``Two Dimensional Black Holes and Planar General
  Relativity'', Class. Quantum Gravity 12, 1081 (1995) 
[arXiv:gr-qc/9407024] 
 
\bibitem{Lemos2}  
J. P. S. Lemos,
``Three Dimensional Black Holes and Cylindrical''
  General Relativity", Phys. Lett. B 352, 46 (1995).    

\bibitem{Lemos3} 
J. P. S. Lemos and V. T. Zanchin, 
``Rotating Charged Black Strings in General Relativity'', 
Phys. Rev. D~{\bf 54}, 3840 (1996)
[arXiv:hep-th/9511188]. 

\bibitem{Huang}
C.-g.~Huang and C.-b~Liang, 
``A torus-like black hole,''  
Phys.~Lett.~A {\bf 201} 27 (1995).  

\bibitem{Aminneborg}
S.~Aminneborg, I.~Bengtsson, S.~Holst, and P.~Peldan, 
``Making Anti-de Sitter Black Holes,'' 
Class.~Quant.~Grav. {\bf 13} 2707 (1996)
[arXiv:gr-qc/9604005].  

\bibitem{brill}
D.~R.~Brill
``Multi-Black-Holes in 3D and 4D anti-de Sitter Spacetimes,'' 
Helv.~Phys.~Acta {\bf 69} 249 (1996)
[arXiv:gr-qc/9608010].    

\bibitem{Mann}
R.~B.~Mann, 
``Pair Production of Topological anti de Sitter Black Holes,'' 
Class.~Quant.~Grav.~{\bf 14} L109 (1997) 
[arXiv:gr-qc/9607071]  

\bibitem{Anderson}
M.~Anderson, P.~T.~Chrusciel, and E.~Delay, 
``Non-trivial, static, geodesically complete, vacuum space-times 
with a negative cosmological constant,''
JHEP {\bf 0210}, 063 (2002)  
[arXiv:gr-qc/0211006].  

%

\bibitem{penrose}
R. Penrose, Riv. Nuovo Cimento 1 (1969) 252.

\bibitem{ChandHartle}
S.~Chandrasekhar and J.~Hartle, Proc.~R.~Soc.~London, 
{\bf A384}, 301 (1982). 

\bibitem{bradysmith}
P.~R.~Brady and J.~D.~Smith, 
``Black hole singularities: a numerical approach,''  
Phys.~Rev.~Lett. {\bf 75} (1995) 1256 
[arXiv:gr-qc/9506067]. 

\bibitem{LeviRoss2003}
T.~S. Levi and S.~F.~Ross, 
``Holography beyond the horizon and cosmic censorship,'' 
Phys.~Rev.~D~{\bf 68} 044005~(2003)
[arXiv:hep-th/0304150].  

\bibitem{Balasubramanian2004}
V.~Balasubramanian and T.~S.~Levi, 
``Beyond the veil: Inner horizon instability and holography,'' 
Phys.~Rev.~D~{\bf 70} 106005 (2004) 
[arXiv:hep-th/0405048].  

\bibitem{horowitzross1997}
G.~T.~Horowitz and S.~F.~Ross, 
``Naked Black Holes,'' 
Phys.~Rev.~D {\bf 56} 2180 (1997)
[arXiv:hep-th/9704058]. 

\bibitem{MannCop}
K.~Copsey and R.~Mann, 
``Pathologies in Asymptotically Lifshitz Spacetimes,'' 
JHEP {\bf 1103}, 039 (2011)
[arXiv:1011.3502[hep-th]]. 

\bibitem{Fernandez}
J.~Fernandez-Gracia and B.~Fiol, ``A no-hair theorem for extremal black branes,"
JHEP {\bf 11}, 054 (2009)
[arXiv:0906.2353 [hep-th]].

\bibitem{Chamblin99}
A.~Chamblin, R.~Emparan, C.~V.~Johnson, and R.~C.~Myers, 
``Holography, Thermodynamics and Fluctuations of Charged AdS Black Holes,'' 
Phys.~Rev.~D~{\bf 60} 104026~(1999)
[arXiv:hep-th/9904197].   

\bibitem{kodama-ishibashi2004}
H.~Kodama and A.~Ishibashi, 
``Master equations for perturbations of generalised static black holes with charge 
in higher dimensions,'' 
Prog.~Theor.~Phys. {\bf 111} 29 (2004).  
[arXiv:hep-th/0308128].  

\bibitem{yoshino}
H.~Yoshino, T.~Ohba, and A.~Tomimatsu, 
``Static black holes with a negative cosmological constant: Deformed 
horizon and anti-de Sitter boundaries,'' 
Phys.~Rev.~D {\bf 69} 124034 (2004)   
[arXiv:gr-qc/0402082]. 

\bibitem{Gowdy}
R.~Gowdy, Ann.~Phys.~{\bf 83}, 203 (1974). 

\bibitem{Moncrief}
B.~Grubisic and V.~Moncrief, 
``Asymptotic Behavior of the $T^3 \times R$ Gowdy Spacetimes,'' 
Phys.~Rev.~D {\bf 47} 2371 (1993)   
[arXiv:gr-qc/9209006].  

\bibitem{Berger}
B.~K.~Berger, D.~Garfinkle, J.~Isenberg, V.~Moncrief, M.~Weaver, 
``The Singularity in Generic Gravitational Collapse Is Spacelike, Local, and Oscillatory,'' 
Mod.~Phys.~Lett.~A {\bf 13} 1565 (1998)   
[arXiv:gr-qc/9805063]. 
 
\bibitem{horowitzross1998}
G.~T.~Horowitz and S.~F.~Ross, 
``Properties of Naked Black Holes,''
Phys.~Rev.~D {\bf 57} 1098 (1998) 
[arXiv:hep-th/9709050]. 

\bibitem{Hawking}
S.W. Hawking and G. F. R. Ellis, The large scale structure of space-time (Cambridge
University Press, Cambridge, 1973).

\bibitem{horowitzway}
G.~T.~Horowitz and Benson Way, ``Lifshitz Singularities,'' 
arXiv:1111.1243~[hep-th]  

\bibitem{Batrachenko:2004fd}
  A.~Batrachenko, J.~T.~Liu, R.~McNees, W.~A.~Sabra, W.~Y.~Wen,
  JHEP {\bf 0505}, 034 (2005)
  [hep-th/0408205].


\end{thebibliography}
\end{document}